\documentclass[a4paper,dvipdfmx]{article}
\makeatletter
\def\@seccntformat#1{\@ifundefined{#1@cntformat}%
   {\csname the#1\endcsname\quad}  
   {\csname #1@cntformat\endcsname}
}
\let\oldappendix\appendix 
\renewcommand\appendix{%
    \oldappendix
    \newcommand{\section@cntformat}{\appendixname~\thesection\quad}
}
\makeatother

\usepackage[dvipdfmx]{graphicx}

\usepackage{cite}
\usepackage[normalem]{ulem}
\usepackage{comment}
\usepackage{url}

\usepackage[usenames,dvipsnames]{xcolor} 

\usepackage{amssymb,amsfonts,amsmath}
\usepackage[top=25truemm,bottom=25truemm,left=20truemm,right=20truemm]{geometry}
\usepackage{bm}
\usepackage{cases}
\usepackage[mathlines]{lineno}

\usepackage{setspace}

\title{Robustness of football passing networks\\against continuous node and link removals}
\bigskip
\author{Genki Ichinose${}^{1*}$, Tomohiro Tsuchiya${}^{1}$, and Shunsuke Watanabe${}^{1}$ 
\ \\
\ \\
${}^{1}$
Department of Mathematical and Systems Engineering, Shizuoka University, \\3-5-1 Johoku, Naka-ku, Hamamatsu, 432-8561, Japan\\
$^*$ Corresponding author (ichinose.genki@shizuoka.ac.jp)}

\begin{document}

\maketitle

\section*{Abstract}
We can construct passing networks when we regard a player as a node and a pass as a link in football games.
Thus, we can analyze the networks by using tools developed in network science.
Among various metrics characterizing a network, centrality metrics have often been used to identify key players in a passing network.
However, a tolerance to players being marked or passes being blocked in a passing network, namely the robustness of the network, has been poorly understood so far.
Because the robustness of a passing network can be connected to the increase of ball possession, it would be deeply related to the outcome of a game.
Here, we developed position-dependent passing networks of 45 matches by 18 teams belonging to the Japan Professional Football League.
Then, nodes or links were continuously removed from the passing networks by two removal methods so that we could evaluate the robustness of these networks against the removals.
The results show that these passing networks commonly contain hubs (key players making passes).
Then, we analyzed the most robust networks in detail and found that their full backs increase the robustness by often invoking a heavier emphasis on attack.
Moreover, we showed that the robustness of the passing networks and the team performance have a positive correlation.

\section*{Keywords}
Football passing networks, network science, robustness, attack, error

\section{Introduction\label{sec:introduction}}
Team sports contain complex interactions with the players in their own and opposing teams and the dynamics excite many audiences.
Team sports such as football (soccer), basketball, rugby, and hockey can be classified into invasion sports because players score by putting a ball (or puck) into their opponent's goal while they also defend their goal against attacks by their opponent \cite{Gudmundsson2017ACMComputSurv}.

In football, each player passes the ball to another player on the team.
Thus, if we consider a player as a node and a pass as a link, we can construct a passing network which allows us to scientifically analyze the network by using various tools developed in network science \cite{Gudmundsson2017ACMComputSurv, Buldu2018FrontPsychol, Buldu2019SciRep, Narizuka2014PhysicaA}.
The ways of constructing passing networks can be generally classified into three types \cite{Gudmundsson2017ACMComputSurv}: (i) player passing networks where a player corresponds to a node and a pass corresponds to a link \cite{Grund2012SocNetworks}, (ii) pitch passing networks where a specific area in the field corresponds to a node instead of a player and the areas are connected by passes (links) \cite{Cintia2015MLDMSAW}, or  (iii) pitch-player passing networks where a player in a certain area at the moment of the pass is a node \cite{Cotta2013JSystSciComplex, Narizuka2014PhysicaA}.

Once a network is constructed, we can apply some metrics developed in network science. See, for example, Ref.~\cite{Buldu2019SciRep} where various metrics have been applied.
Previous studies have focused on which players are important in passing networks.
In those analyses, the most used metric was centrality.
There are various kinds of centrality.
In football analyses, degree centrality \cite{Grund2012SocNetworks}, betweenness centrality \cite{Pena2012ProcEPSC}, flow centrality \cite{Duch2010PLoSONE}, closeness centrality \cite{Pena2012ProcEPSC}, and eigenvector centrality \cite{Cotta2013JSystSciComplex} have been applied.
For example, Grund focused on degree centrality and analyzed a huge amount of passes in English Premier League games.
He found that concentrating too much on passes to certain players results in decreasing team performance \cite{Grund2012SocNetworks}.
Pe\~{n}a and Touchette focused on betweeness centrality \cite{Pena2012ProcEPSC}.
Betweeness centrality measures how the flow of the ball between players depends on certain players.
Namely, it measures how many passes are conducted through those players.
Thus, we can consider that it reflects the impact of removing those players.
They argued that it is important that the betweeness centrality of each player should not be biased to specific players from a tactical point of view.

Apart from centrality, clustering coefficient is also used as one of the metrics \cite{Pena2012ProcEPSC}.
This metric calculates the degree of triangles for all the neighbors for a certain player.
Thus, it reflects the contribution of the player against the local robustness of the passing network.
It is also known that the coordination of three professional players which form a triangle is far more elaborated than that of beginners in a ball passing situation \cite{Yokoyama2018PhysRevE}.

However, in those traditional analyses of football passing networks, the robustness of the networks has been poorly understood where players or passes are subject to continuous removals.
Because the robustness of a passing network can be directly connected to the increase of ball possession, it would be related to the outcome of a game.
In network science, with the pioneering work by Albert et al. \cite{Albert2000Nature}, the robustness of networks against continuous node removals has been successively addressed \cite{Albert2002RevModPhys, Holme2002PhysRevE, Strogatz2001Nature, Newman2003SIAMRev, Goh2002PhysRevLett, Paul2004EurPhysJB}.
Therefore, we aim to identify the characteristics of passing networks in football by applying those discoveries.

In this paper, we constructed position-dependent passing networks from coordinate data of passes in 45 matches of the Japan Professional Football League (J League).
Then, we analyzed the robustness of the networks against the two types of continuous node or link removals.
As a result, we found that all of those passing networks have hubs.
We further conducted the analysis of the most robust team.

\section{Methods}
\subsection{Dataset}
We use the data of 45 matches of the J1 league (the top division of J League) as the dataset provided by DataStadium Inc., Japan.
DataStadium has a contract with J League to collect and sell data.
We use the dataset for this research under the permission by DataStadium.
The dataset is composed of five matches per team for all 18 teams in the second stage of 2018 (from August 10 to September 2), that is, $(18 \times 5)/2=45$.
We label the 18 teams ``Kawasaki'', ``Hiroshima'', ``Kashima'', ``Sapporo'', ``Urawa'', ``Tokyo'', ``C Osaka'', ``Shimizu'', ``G Osaka'', ``Kobe'', ``Sendai'', ``Yokohama'', ``Shonan'', ``Tosu'', ``Nagoya'', ``Iwata'', ``Kashiwa'', ``Nagasaki'', respectively.
Note that we sorted the teams in ascending order based on the annual ranking in the 2018 season, namely, ``Kawasaki'' was the champion and ``Nagasaki'' was the lowest in that season.

We used NetworkX for the visualization and analyses of networks. NetworkX is a Python library.

\subsection{Construction of passing networks}
We constructed a passing network from coordinate data of incoming and outgoing passes and players' numbers where a player corresponds to a node and a pass corresponds to a link.
We only used the data of successful passes and excluded passes stolen by the opponent team or plays out of bounds.
We included throw-ins and set-pieces in the data.
The obtained data was 36,070 lines in total.
The sample of the data is shown in Table \ref{pass-data}.
Each line represents pass sender's and receiver's IDs with their locations (X-coordinates and Y-coordinates).
From this data, we constructed passing networks.

\begin{table}[htbp]
  \centering
  \begin{tabular}{|c|c|c|c|c|c|} \hline
    \multicolumn{3}{|c|}{Pass sender} &  \multicolumn{3}{c|}{Pass receiver} \\ \hline
    Player ID & X-coordinate & Y-coordinate & Player ID & X-coordinate & Y-coordinate \\ \hline
    1200017 & $-6.5$ & 85.5 & 1400359 & $-69$ & 70 \\
    1400359 & $-69$ & 70 & 1100134 & $-110$ & 74.5 \\
    1301053 & $-139.5$ & 104 & 1000646 & $-113.5$ & 85.5 \\
    - & - & - & - & - & - \\ \hline    
  \end{tabular}
  \caption{Sample of football passing data. Only three lines are shown although the data contains 36,070 lines in total.}
  \label{pass-data}
\end{table}

\begin{figure}[h]
	\centering
	\includegraphics[width=\textwidth]{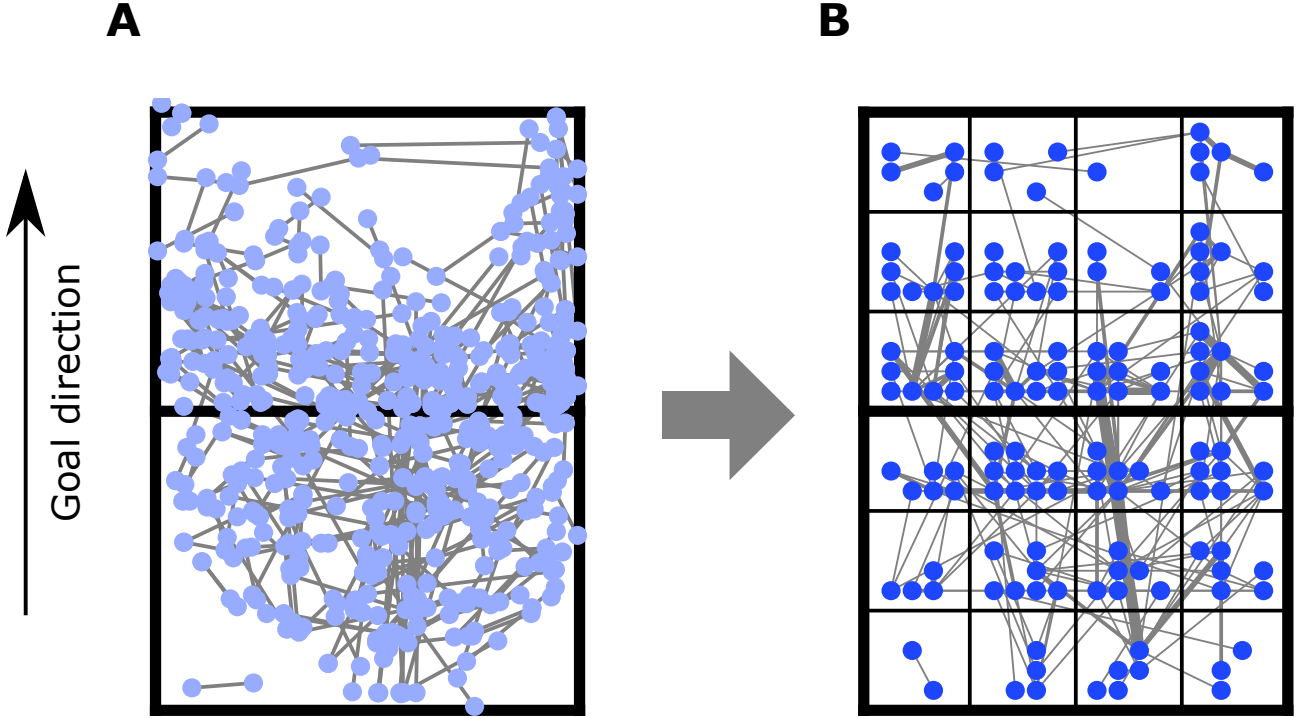}
	\caption{Passing network created from the data of Nagoya in Nagoya vs.~Yokohama on August 15, 2018. (A) Original passing network. (B) Position-dependent passing network transformed from (A). Multiple links are allowed for a pair of nodes.}
	\label{fig:pass_networks}
\end{figure}

We constructed a passing network per team in a match.
An example is shown in Fig.~\ref{fig:pass_networks}.
This is a match between Nagoya versus Yokohama and the network corresponds to Nagoya.
For Nagoya, the direction of offense is upward.
Figure \ref{fig:pass_networks}(A) corresponds to the original passing network.
In the network, we create a node from a coordinate where a player passes the ball (start point) to the other coordinate where another player receives the ball (end point).

In the Fig.~\ref{fig:pass_networks}(A) style, because each location is regarded as a different node, it is unclear which players collect passes.
Thus, for our analyses, we employ position-dependent networks where the field is divided into some areas depending on the locations \cite{Cotta2013JSystSciComplex, Narizuka2014PhysicaA, Narizuka2015JPhysSocJpn}.
We can consider several different ways to divide the field. Here, we divided it into $4 \times 6=24$ and constructed networks.
In our position-dependent networks, we regard a player with different locations in the same area as the same (one node).
In contrast, we distinguish passes with different locations in the same area as different passes.
Thus, the position-dependent networks allow multiple links.

Figure \ref{fig:pass_networks}(B) is the position-dependent passing network transformed from the original passing network (Fig.~\ref{fig:pass_networks}(A)).
In Fig.~\ref{fig:pass_networks}(A), a player with different locations in the same area is regarded as different nodes because their coordinates are different.
In Fig.~\ref{fig:pass_networks}(B), the player is regarded as one node.
However, passes in the same area are not integrated into one pass.
Thus, the network is a weighted network which reflects the number of passes.
The line thicknesses represent the weights.
In Fig.~\ref{fig:pass_networks}(B), the players (nodes) are aligned from the bottom left to the top right diagonally, and they move horizontally from left to right based on the players' numbers in each line.
That is, in the figure, the exact locations of passes are not reflected.
In the following, we analyze these position-dependent passing networks.

\subsection{Network models for comparison}
To identify the characteristics of the robustness of passing networks in football, we constructed two-type model-based networks for comparison. 
These two networks are Exponential (E) network and Scale-free (SF) network.
These were analyzed in the network tolerance study \cite{Albert2000Nature}.
We also adopted these two networks as the representatives although we can consider other different types of networks.
Those two networks can be considered extreme in the sense that E networks are random and SF networks are centralized.
Obviously, football passing networks are neither random nor scale-free. 
However, those two types of well-known networks are still useful for comparison with our football passing networks to extract the characteristic properties of the passing networks.

In the E networks, most of the nodes have similar degrees and the nodes are randomly connected.
We generated these networks by using Er\H{o}s-R\'enyi's algorithm \cite{Erdos1960PublMathInstHungAcadSci}.
In the original algorithm, all pairs are selected once and each pair is connected with a certain probability $p$.
Here, to compare them with football passing networks, we randomly select a pair and connect the pair with $p$.
This is repeated for the number of pairs instead of selecting all pairs once.
Thus, the generated E networks allow multiple links.

On the other hand, in the SF networks, most of the nodes have a few links while a few nodes, called hubs, gather many links.
The degree distributions show a power-law.
We generated these networks by using Barab\'asi-Albert algorithm \cite{Barabasi1999Science}.
Initially, a network starts from a few nodes which are connected by links.
Then, a new node is repeatedly introduced. Every new node is connected to existing nodes in proportion to the degree of existing nodes.
This mechanism is called a preferential attachment.
In the original algorithm, multiple links are not allowed but here we allow them to compare with football passing networks.
It means that when a new node is introduced, it can connect to the same node multiple times.

Based on those algorithms, we generated 100 networks for each. The size of the networks is $N=150$ and the average degree is $\langle k \rangle \simeq 8$.
These parameter values were used because they are suitable to compare with football passing networks. The values of the passing networks are shown later.

\subsection{Continuous node or link removals}
We analyze the robustness of passing networks of all J1 teams by conducting continuous node or link removals on the networks (e.g., Fig.~\ref{fig:pass_networks}(B)).
We compare the robustness of the networks with that of E and SF networks by removing nodes or links in the two types of networks as well.

For node removals, we employ the two ways used in Albert's model \cite{Albert2000Nature}.
The first one is called ``error'' where a node is randomly selected and removed, one after another.
The other one is called ``attack'' where the largest hub is selected and removed, one after another.
For link removals, we also employ ``error'' and ``attack.''
In the error, a pair of two nodes are randomly selected.
In the attack, the pair of nodes which has the maximum links is selected.
In both cases, all links between the selected pair of nodes are removed.

Some possible interpretations of the continuous node or link removals in football games are as follows.
Continuous node removals correspond to repeated man-marks for specific players when focusing on plays or repeated zone defenses when focusing on positions.
By those defenses, players become less functional.
On the other hand, continuous link removals may be easier to interpret than node removals.
The link removals correspond to the situation that passes between a pair of two players are blocked or lanes are closed one after another by the opponent's team.

We model these situations by continuous node or link removals and analyze the robustness of networks.

\subsection{Robustness measurement}
We use three measures, the diameter of a network, $d$, the algebraic connectivity (the second-smallest eigenvalue of the Laplacian matrix of a network), $\lambda_2$, and the relative size of the largest cluster, $S$, to quantify the robustness of networks against the number of removed nodes, $n_{\rm R}$ \cite{Albert2000Nature}, or links, $l_{\rm R}$.

A network diameter denotes the longest path among the shortest paths between any pair of nodes.
In general, when a hub node is removed, the diameter becomes longer because the removal makes many of the shortest paths vanish.
An algebraic connectivity reflects how well connected the overall network is.
If the value of a network is large, it implies the network is well connected.
The value becomes zero when a network splits into two components.
The largest cluster implies the largest connected subgraph in a network.
We use the relative size of the largest cluster. The value of $S$ starts from 1 and it decreases as nodes are removed.
We evaluate the robustness of passing networks by analyzing the change of the three measures $d$, $\lambda_2$, and $S$ against continuous node or link removals.

Here, we explain how the process of removing nodes or links works to determine the robustness of a network by the three measurements.
We continuously conduct node removals or link removals for each passing network.
A network is regarded as robust when the diameter $d$ remains low even after nodes or links are largely removed because the shortest paths between nodes are not lost.
Also, when the algebraic connectivity $\lambda_2$ remains high even after nodes or links are largely removed because it implies that the network is well connected as a whole.
Finally, when the relative size of the largest cluster $S$ remains high even after nodes or links are largely removed because it implies that the network is not split into multiple sub-networks.

\section{Results}
\subsection{Passing network characteristics}
First, we summarize the characteristics of the position-dependent passing networks of J1 teams typically represented by Fig.~\ref{fig:pass_networks}(B) in Table \ref{table:network_char}.
In Table \ref{table:network_char}, the teams are sorted in descending order of the number of weighted links (passes).
Here, we only show the top three and the bottom three teams in all 18 teams.
The data of all teams are provided in Supplementary Material (Table S1).
The values are averaged over five matches.
Obviously, the number of passes of Kawasaki is much higher than the other teams.
Kawasaki is known for possessing the ball by passing to one another \cite{KawasakiURL}.
Thus, our results supported this fact.

\begin{table}[hbtp]
\centering
\caption{Characteristics of the position-dependent passing networks of J1 teams sorted by the number of weighted links (passes) in descending order. Only the top three and bottom three teams are shown.}
\begin{tabular}{lcrrr}
\hline
                 & Team     & \multicolumn{1}{c}{\# of links} & \multicolumn{1}{c}{\# of nodes} & \multicolumn{1}{c}{$\left<k\right>$} \rule[-10pt]{0pt}{25pt} \\ \hline
                 & Kawasaki & 719.8                             & 164.4 & 8.76 \rule[-10pt]{0pt}{25pt} \\
Top 3 & Kobe & 505.6                           & 142.2 & 7.11 \rule[-10pt]{0pt}{20pt} \\
                 & Sapporo  & 452.8                           & 154.0 & 5.88 \rule[-10pt]{0pt}{20pt} \\ \hline
                 & Shonan  &  313.2                          & 147.2 & 4.26 \rule[-10pt]{0pt}{25pt} \\
Bottom 3  & Shimizu  & 296.0                           & 137.4 & 4.31 \rule[-10pt]{0pt}{20pt} \\
                 & Tosu     & 264.0                           & 132.6 & 3.98 \rule[-10pt]{0pt}{20pt} \\ \hline
\end{tabular}
\label{table:network_char}
\end{table}

\subsection{Passing network robustness}
\subsubsection{Diameter change}
\begin{figure}[h]
	\centering
	\includegraphics[width=\textwidth]{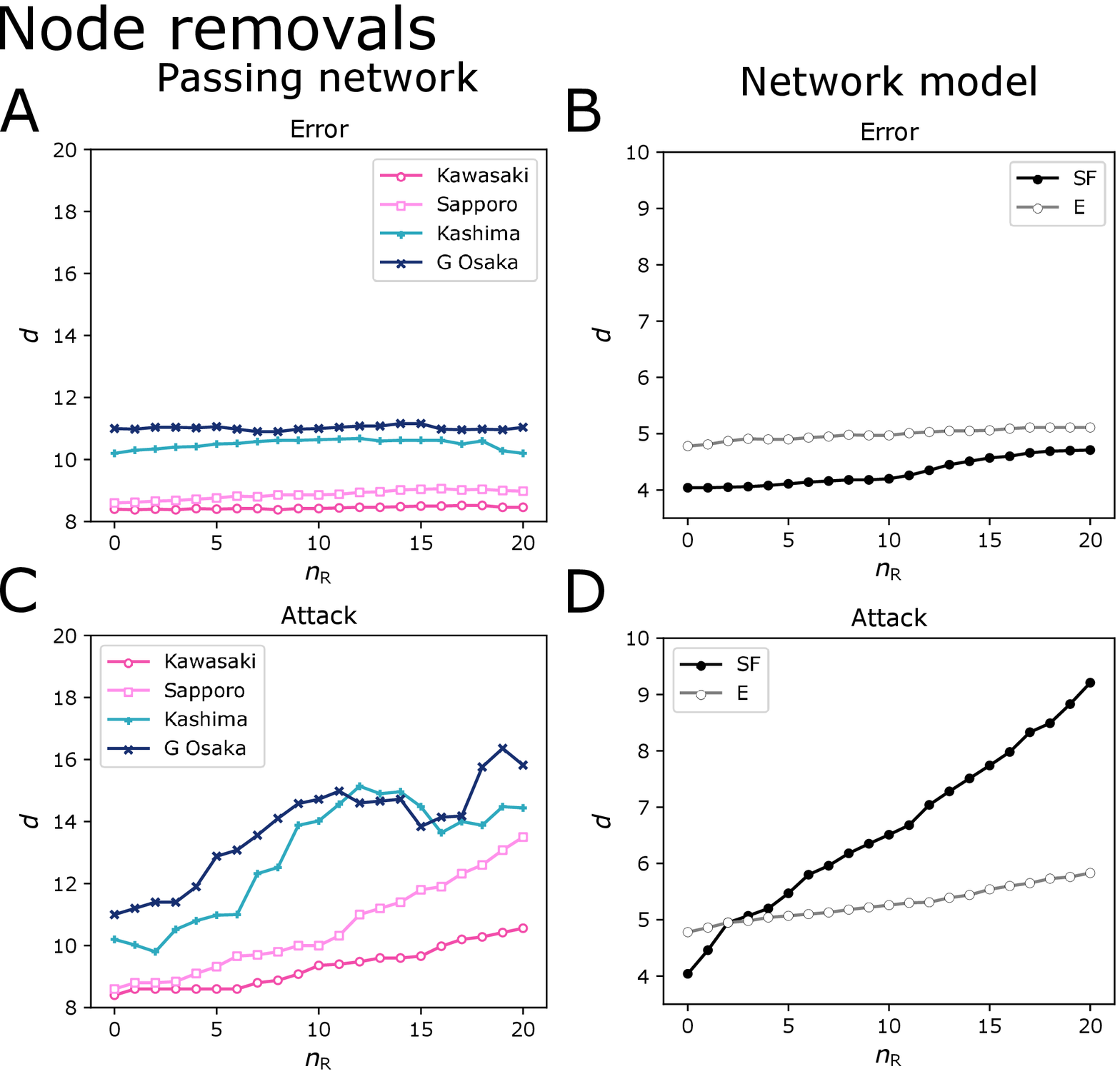}
	\caption{Change of the diameter $d$ against node removals $n_R$ in errors (A and B) and attacks (C and D). (A and C) are the cases of football passing networks. Five matches are averaged for each line.  (B and D) are the cases of the E and SF networks. One-hundred realizations are averaged for each line. Note that the scales between the passing networks and the network models are different because the former has a spatial limitation as explained in the text.}
	\label{fig:diameter_N}
\end{figure}

Figure \ref{fig:diameter_N} shows the changes of the diameter $d$ for football passing networks (Figs.~\ref{fig:diameter_N}(A) and (C)) and the two network models (Figs.~\ref{fig:diameter_N}(B) and (D)) against node removals.
Errors (random node removals) correspond to (A) and (B) while attacks (hub node removals) correspond to (C) and (D).
We only show the biggest two and smallest two teams for $d$ at $n_{\rm R}=6$ against errors in Fig.~\ref{fig:diameter_N}(A) and (C).
The biggest two teams are G Osaka and Kashima from the top and the smallest two teams are Kawasaki and Sapporo from the bottom in Fig.~\ref{fig:diameter_N}.

\begin{figure}[h]
	\centering
	\includegraphics[width=\textwidth]{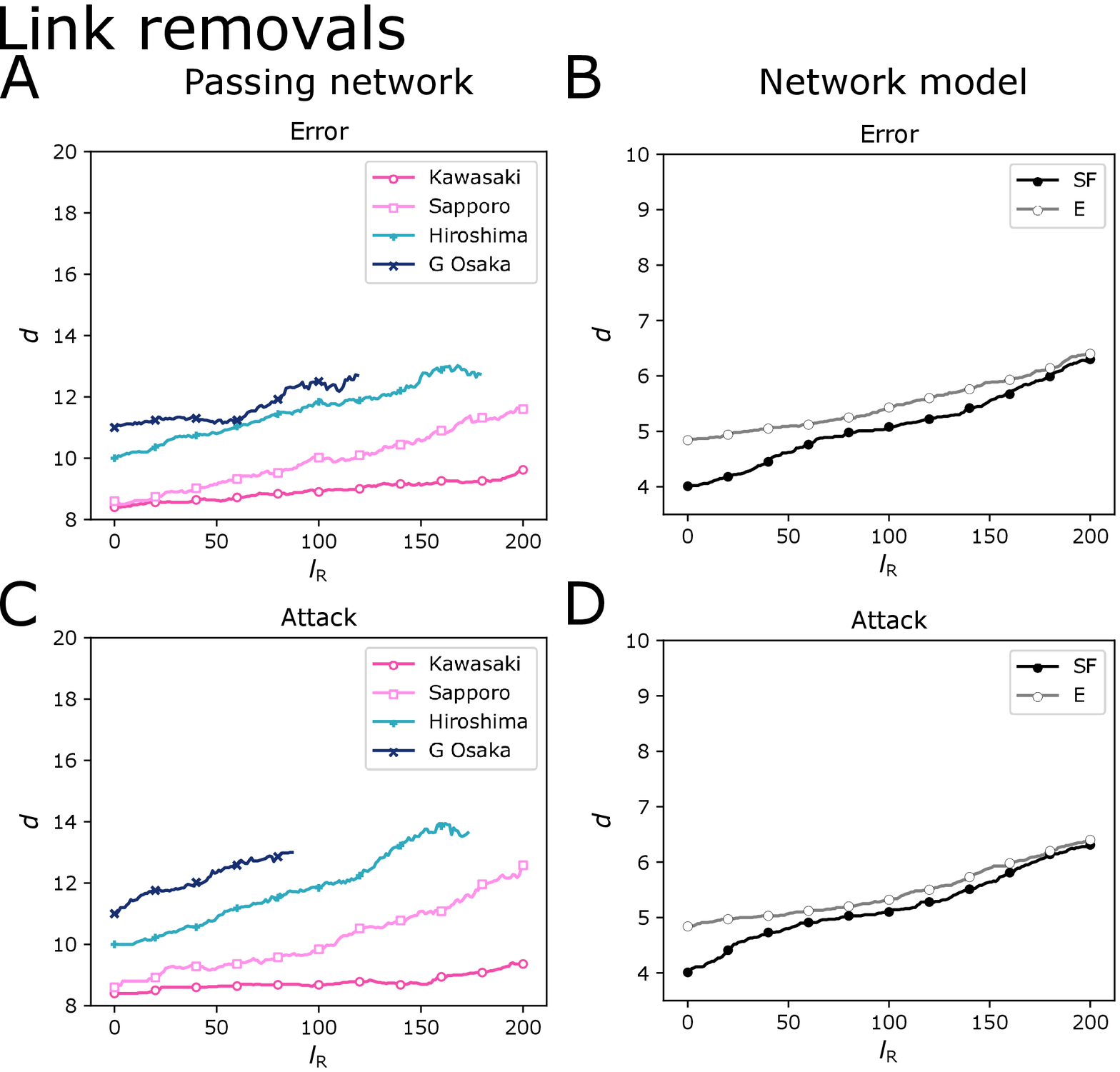}
	\caption{Change of the diameter $d$ against link removals $l_R$ in errors (A and B) and attacks (C and D). The same settings as Fig.~\ref{fig:diameter_N} are used.}
	\label{fig:diameter_L}
\end{figure}

First, we focus on the change of $d$ against errors.
Figure \ref{fig:diameter_N}(A) shows the case of the football passing networks.
The value of $d$ in the four teams does not change even if $n_{\rm R}$ is increased, which suggests that football passing networks are robust against errors.
On the other hand, the values of $d$ in Kawasaki and Sapporo are smaller than those in G Osaka and Kashima.

We compare the passing networks with the two network models.
When $n_{\rm R}$ is increased, the value of $d$ does not change in the E networks while that value slightly increases in the SF networks.
In the SF networks, when a hub is removed from a network, it greatly increases $d$.
This situation sometimes happens because we adopted the small size of network ($N=150$).
(This situation does not happen when the network size is large as shown in Albert et al. \cite{Albert2000Nature}).
By comparing Fig.~\ref{fig:diameter_N}(A) with Fig.~\ref{fig:diameter_N}(B), we found that $d$ in the passing networks ($8 \leq d \leq 11$) is two to three times larger than that in the E and SF networks ($4 \leq d \leq 5$).
The position-dependent passing networks in football have a spatial limitation.
In the networks, short passes are more often observed than long passes.
In other words, passes to the next area in the field tend to be larger.
In contrast, there is no spatial limitation in the E and SF networks.
Thus, the values of $d$ in the passing networks are much larger than the ones in the network models.

Next, we focus on the change of $d$ against attacks.
Figure \ref{fig:diameter_N}(C) shows the case of the football passing networks.
The value of $d$ in the four teams increases as $n_{\rm R}$ is increased, which suggests that football passing networks are vulnerable to attacks.
However, the amount of change in Kawasaki is small even if $n_{\rm R}$ is increased.
It means that Kawasaki's networks are tolerant against attacks to a certain extent.

We compare the passing networks with the two network models.
The diameter of the E networks was slightly increased against attacks (Fig.~\ref{fig:diameter_N}(D)).
Basically, there is not much difference among the degrees of nodes in the E networks.
However, because the network size is small ($N=150$), it generates the tendency that some nodes have relatively higher degrees while some other nodes have relatively smaller degrees.
Thus, when the higher degree nodes are removed by attacks, the value of $d$ becomes larger even in the E networks.
In contrast, the value of $d$ in the SF networks is rapidly increased against attacks because shortest paths vanish due to the hub node removals.
Thus, the SF networks are extremely vulnerable to attacks.
As shown in Fig.~\ref{fig:diameter_N}(C), we observed that the values of $d$ in the passing networks increase as $n_{\rm R}$ becomes larger although the change in Kawasaki is small.
This is because the passing networks are characterized by the existence of some hub nodes (key players making passes).

We see the same measurement $d$ against link removals in Fig.~\ref{fig:diameter_L}.
The same labeling with Fig.~\ref{fig:diameter_N} is used for Fig.~\ref{fig:diameter_L}.
We pick the biggest two and smallest two teams for $d$ at $l_{\rm R}=6$ against errors in Fig.~\ref{fig:diameter_L}(A) and (C).
The difference with Fig.~\ref{fig:diameter_N} was that the second biggest team was Hiroshima instead of Kashima.
In (A) and (C), we exclude the data when the decrease of $d$ which is larger than 0.05 continues twice in a row because it basically suggests that the network is split into two or more components.

Compared to the node removals (Fig.~\ref{fig:diameter_N}) with the link removals (Fig.~\ref{fig:diameter_L}), $d$ becomes larger as $l_R$ increases even in the case of errors.
This is because all links between two selected nodes are removed in the case of link removals.
It contributes to the increase of $d$ because paths tend to disappear due to the link removals.
Thus, in the case of link removals, the difference between errors and attacks is diluted although the increase in attacks is slightly bigger than that in errors, which makes it difficult to characterize passing networks as $d$ of link removals.
However, $d$ in Kawasaki's network remains low against errors and attacks, which suggests that Kawasaki's network is robust.

In summary, we found that the passing networks were well characterized by node removals rather than link removals for $d$.
In the node removals, the passing networks were robust against errors but vulnerable to attacks, suggesting that the passing networks have some hubs.
We also found that the smallest change of $d$ was Kawasaki, thus Kawasaki was most robust against continuous node and link removals in the passing networks.

\subsubsection{Algebraic connectivity change}
\begin{figure}[h]
	\centering
	\includegraphics[width=\textwidth]{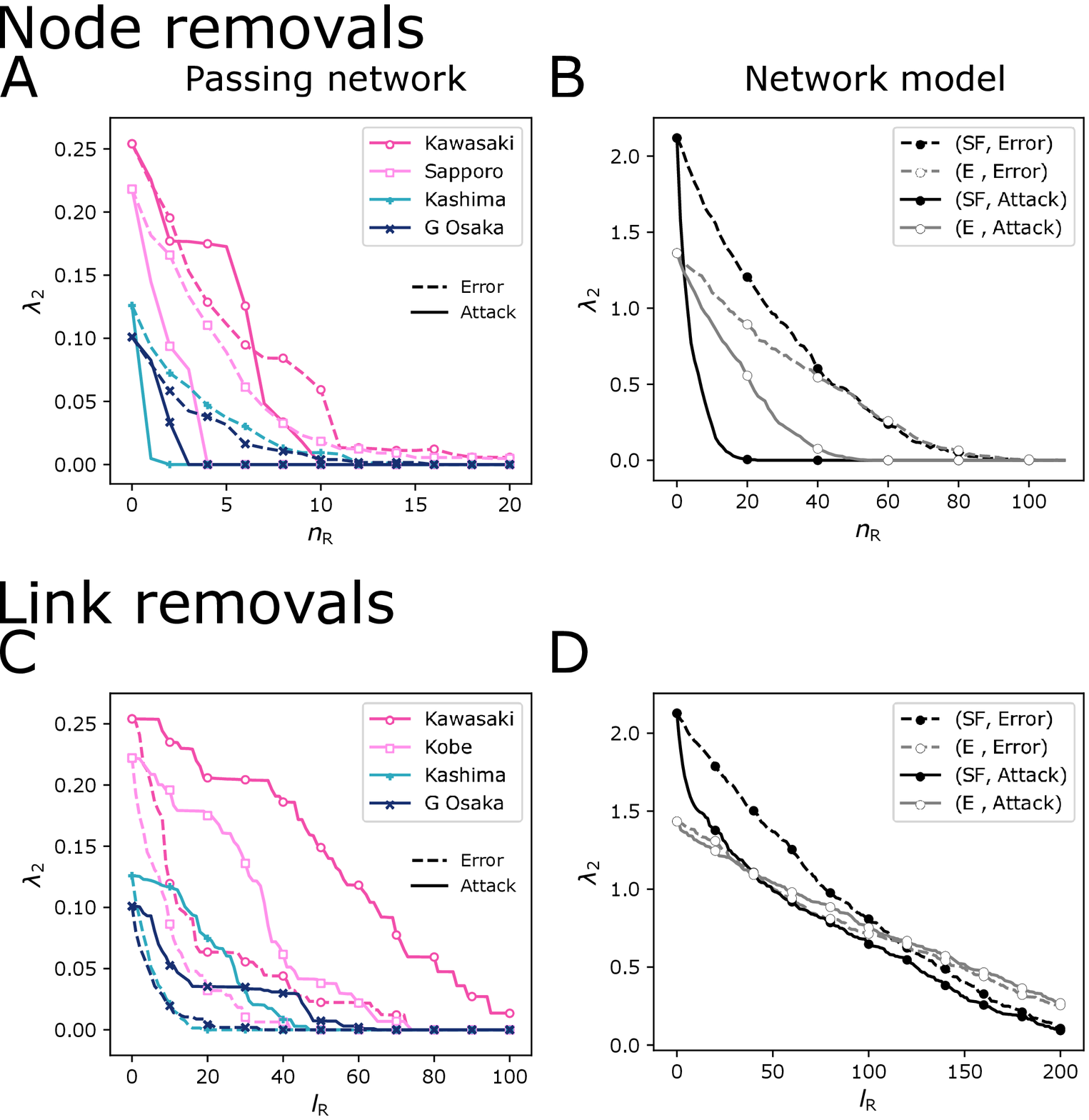}
	\caption{Change of the algebraic connectivity $\lambda_2$. (A) and (C) are the cases of football passing networks against $n_R$ and $l_R$, respectively. Five matches are averaged for each line.  (B) and (D) are the cases of the E and SF networks against $n_R$ and $l_R$, respectively. One-hundred realizations are averaged for each line.}
	\label{fig:laplacian}
\end{figure}

Next, we focus on the change of the algebraic connectivity $\lambda_2$ (Fig.~\ref{fig:laplacian}).
Algebraic connectivity is represented by $\lambda_2$ which is the second-smallest eigenvalue of the Laplacian matrix of a network.
If the value is large, it implies that the network is well connected.
When a network is split into two components, $\lambda_2$ becomes zero.

Figures \ref{fig:laplacian}(A) and (B) represent the case of node removals ($n_R$) where (A) corresponds to passing networks while (B) is network models.
For the passing networks, we only show the biggest two and smallest two teams for $\lambda_2$ at $n_{\rm R}=1$ against errors.
The biggest two teams are Kawasaki and Sapporo from the top and the smallest two teams are G Osaka and Kashima from the bottom.

In both passing networks (A) and network models (B), the values approach zero as $n_R$ becomes larger.
Basically, in both networks, the decreases of $\lambda_2$ in the case of attacks is larger than that in the case of errors.
This is because when hubs are removed by attacks, it greatly decreases the connectivity of a network.
When we focus on the difference between attacks and errors in network models (B), the SF networks are bigger than the E networks.
Because hubs are connected from many other nodes in the SF networks, the effect of hub removals tends to be bigger in the SF networks compared to the E networks.
In passing networks (A), the difference between attacks and errors is large because passing networks contain some hubs.
Note that the value of $\lambda_2$ in network models (B) is much higher than that in passing networks (A).
Passing networks are locally connected due to the position-dependent characteristic, which makes the connectivity throughout the networks weak.

Figures \ref{fig:laplacian}(C) and (D) represent the case of link removals ($l_R$) where (C) corresponds to passing networks while (D) is network models.
For (C), we only show the biggest two and smallest two teams for $\lambda_2$ at $l_{\rm R}=1$ against errors.
The biggest two teams are Kawasaki and Kobe from the top and the smallest two teams are G Osaka and Kashima from the bottom.
The difference with (A) was that the second biggest team was Kobe instead of Sapporo.

The biggest difference between (A) and (C) is that $\lambda_2$ in the case of attacks are higher than that in the case of errors in the link removals (C).
In this case, links around a hub player (key player regarding passes) tend to be selected.
Let's label the key player X and assume that the maximum number of links exists between X and another player Y.
Then, in attacks, the first removal is conducted for all the links between X and Y.
If the second largest number of links were to exist between X and another player Z, then the second attack will be conducted for all the links between X and Z.
In this way, links which belong to the neighbors of a hub player tend to be selected for removal in the case of attacks.
However, even if these links are removed, these removals do not affect the connectivity of the whole network much because these removals are conducted only between the hub node and the neighbors.
In contrast, in the case of errors, links between a randomly selected pair are removed.
These removals affect the connectivity of the whole network because it may split the network.
This is why $\lambda_2$ in the case of attacks is higher than that in the case of errors.

In network models (D), the removal of one pair which has the maximum number of links affects the network connectivity because almost all links are concentrated on only a few nodes. 
As a result, we do not see any similar properties between (C) and (D).

Therefore, we conclude that it is better to use $n_R$ than $l_R$ when we compare passing networks with network models.
When hub nodes (key players) in passing networks are removed, it greatly affects the whole connectivity of the networks.

\subsubsection{Largest cluster change}
\begin{figure}[h]
	\centering
	\includegraphics[width=\textwidth]{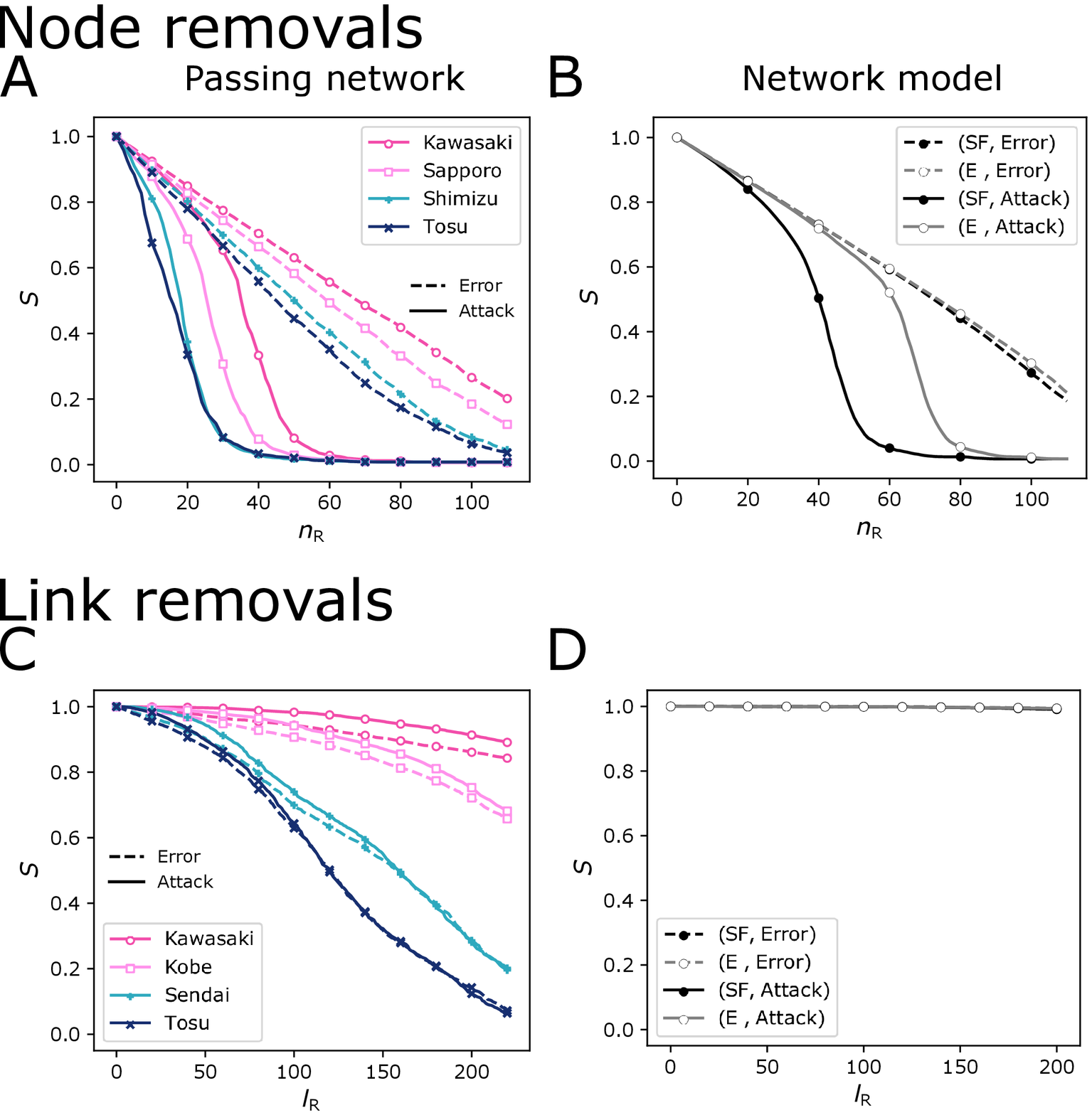}
	\caption{Change of the relative sizes of the largest cluster $S$. (A) and (C) are the cases of football passing networks against $n_R$ and $l_R$, respectively. Five matches are averaged for each line.  (B) and (D) are the cases of the E and SF networks against $n_R$ and $l_R$, respectively. One-hundred realizations are averaged for each line.}
	\label{fig:cluster}
\end{figure}

Finally, we focus on the change of the relative size of the largest cluster $S$ (Fig.~\ref{fig:cluster}).
Figure \ref{fig:cluster}(A) shows the case of the passing networks against node removals ($n_R$).
We only show the biggest two and smallest two teams for $S$ at $n_{\rm R}=80$ against errors.
The values of $S$ in all four teams linearly decrease as $n_{\rm R}$ becomes larger in the case of errors (random node removals), which means that each network is connected as a whole.
On the other hand, the values of $S$ in all four teams suddenly fall as $n_{\rm R}$ becomes larger in the case of attacks.
The reason is that when hub players are removed by attacks (hub node removals), the networks become disconnected and divided into quite small subgraphs.

We compare the passing networks with the two network models against node removals.
We see the changes of $S$ in the E and SF networks (Fig.~\ref{fig:cluster}(B)).
In general, in the E networks, it has been shown that there is no difference against errors and attacks (See Fig.~3a in Albert et al.~\cite{Albert2000Nature}) if the network size is large (e.g., $N=10,000$).
Here, we observed the difference between errors and attacks in the E networks because there are some differences in the degrees among nodes due to the small network size ($N=150$).
In the SF networks, the value of $S$ suddenly falls against attacks but linearly decreases against errors.
Thus, the impact of removing hubs is large in the SF networks.
In Fig.~\ref{fig:cluster}(A), we found that $S$ nosedives in the passing networks, although the bottom two teams (Shimizu and Tosu) are more intense than the SF networks.
These results also suggest that the passing networks have some hubs (key players making passes).

Figures \ref{fig:cluster}(C) and (D) show the same measurement $S$ against link removals ($l_R$).
The same as with Fig.~\ref{fig:laplacian}(C), $S$ in attacks is larger than that in errors.
For the same reason as Fig.~\ref{fig:laplacian}(C), links which belong to the neighbors of a hub player tend to become the targets for removals.
Thus, the largest cluster of a network remains because those removals barely split the network.
There is quite little difference among the results in the two network models (D).
In these networks, each network is well connected as a whole.
Thus, those networks are not affected by the link removals unless the number of removals is much larger.

We again see the characteristics of passing networks are well captured by node removals rather than link removals.
We also found that Kawasaki's network was most robust because the largest cluster of the network decreased to be the slowest against attacks.

\subsection{Detailed analysis of Kawasaki passing network}
\begin{figure}[h]
	\centering
	\includegraphics[width=100mm]{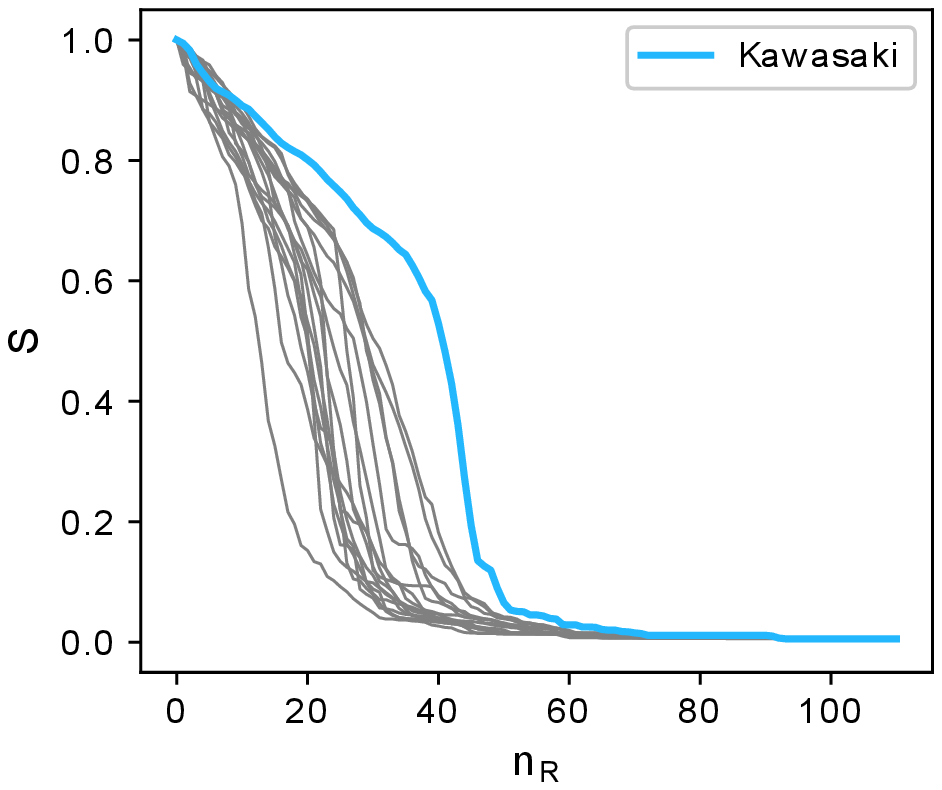}
	\caption{Change of the relative size of the largest clusters $S$ against attacks for all 18 teams.}
	\label{fig:kawasaki_cluster}
\end{figure}

The three analyses above revealed that Kawasaki was the most robust in all 18 teams.
Figure \ref{fig:kawasaki_cluster} shows the changes of the largest cluster $S$ against $n_R$ in all teams.
Obviously, Kawasaki was the most robust by far.
Thus, we focus on Kawasaki's network in detail.

\begin{figure}[h]
	\centering
	\includegraphics[width=150mm]{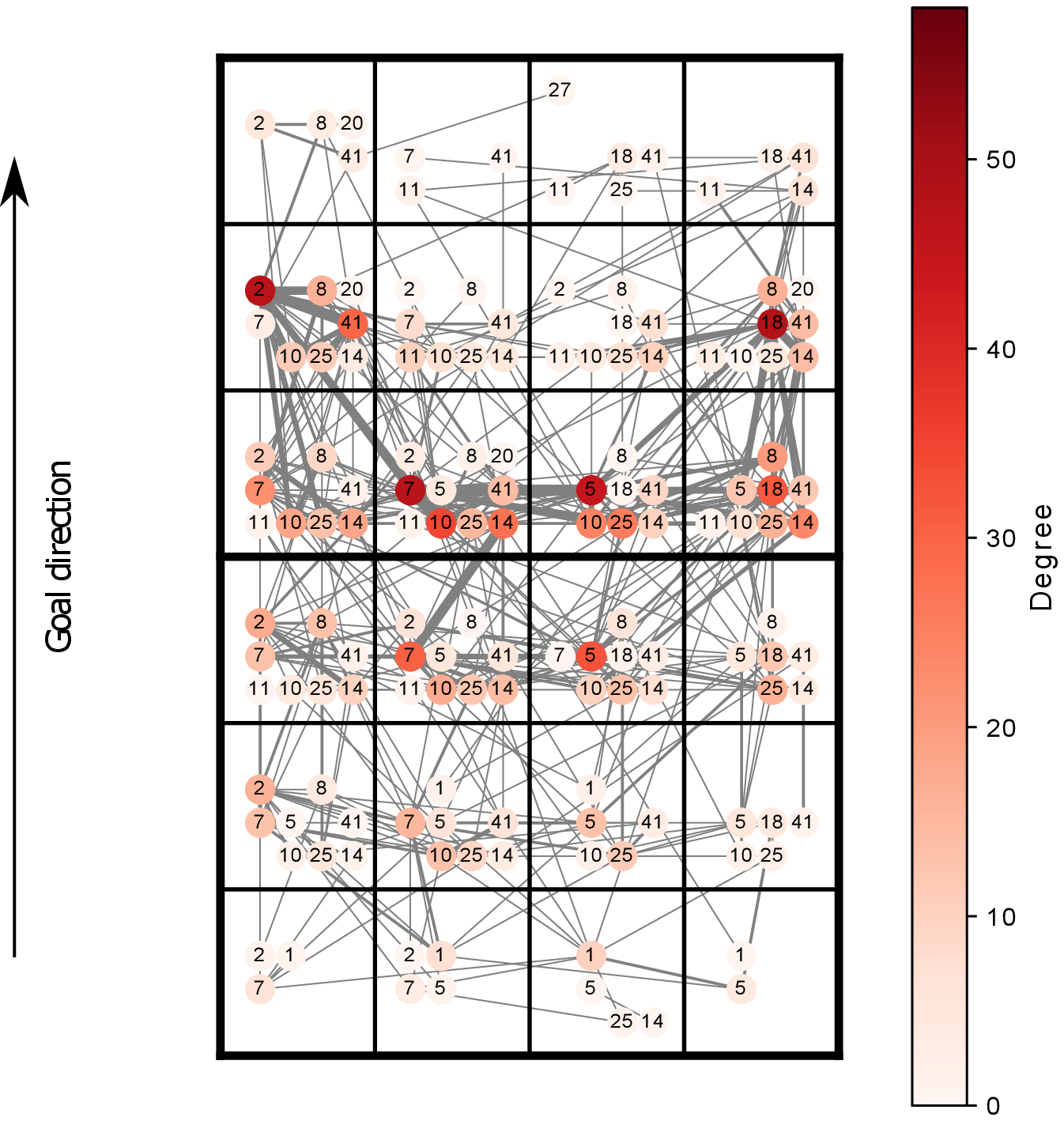}
	\caption{Kawasaki's position-dependent passing network. Kawasaki vs. Hiroshima on August 19, 2018. The node color deepens when the degree of the node is high.}
	\label{fig:pass_network_kawasaki}
\end{figure}

We show Kawasaki's typical network (Fig.~\ref{fig:pass_network_kawasaki}).
In the figure, players' numbers are displayed on the nodes.
The players (nodes) are aligned from the bottom left to the top right diagonally as the same style Fig.~\ref{fig:pass_networks}.
As a player touch the ball earlier, the player appears faster.
Namely, a player who touched the ball fastest corresponds to the bottom left node.
The colors deepen in proportion to the degree (the number of passes).

In the left area, Noborizato (number 2); in the center area, Oshima (number 10) and Morita (number 25); in the right area Ienaga (number 41) and Elsinho (number 18) have deep colors.
It means that those players are hubs which collect passes.
Here, Noborizato (number 2) and Elsinho (number 18) are full backs but they are often observed in the opponent areas, which means they actively assist attacking the opponent's goal.
Also, Ienaga (number 41) is basically a hub on the right side but he is also observed in many other areas.
Thus, he is involved in passing the ball in broad areas.

In many football teams, offensive midfielders and defensive midfielders play the central role of passing the ball.
Not only those central players but also some other players are the hubs of passes in Kawasaki's network.
This signature contributes to make Kawasaki's passing networks robust.

\begin{figure}[h]
	\centering
	\includegraphics[width=150mm]{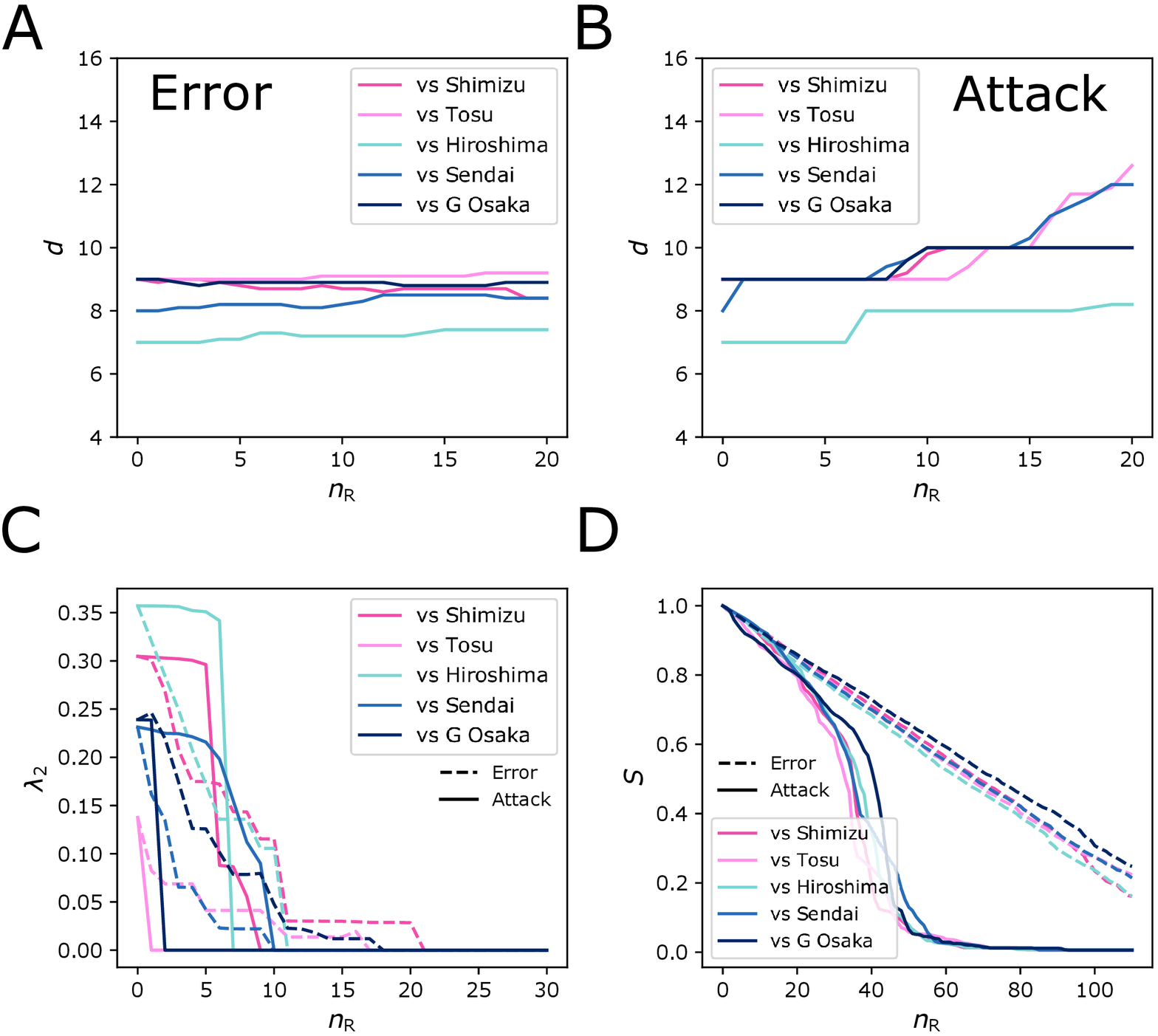}
	\caption{Five individual matches of Kawasaki. (A) Diameter change against errors, (B) Diameter change against attacks, (C) Algebraic connectivity change, and (D) Largest cluster change.}
	\label{fig:each_match_kawasaki}
\end{figure}

One thing that needs to be paid attention to is that we aggregated the data and used the average values for the analyses.
For example, Kawasaki may show good passing effectiveness when playing against some teams but may not when playing against other teams.
To elucidate this question, we separated the aggregated data into five individual matches (Fig.~\ref{fig:each_match_kawasaki}).
Only the node removal cases are shown because they accurately capture the robustness of passing networks rather than the case of the link removals.
We see variations among the matches to some extent for the three measures.
This result suggests that Kawasaki's passing networks change depending on the opponent's teams.
Especially, Kawasaki's network is most vulnerable (large $d$, small $\lambda_2$, and small $S$ as a whole) to 
Tosu.
The win-loss record of Kawasaki versus Tosu for that match (August 15, 2018) was a draw (0--0).
In this way, we must pay attention to the data at the individual match level because there are variations to some extent.
However, we also find that Kawasaki's networks are robust against the five teams (Fig.~\ref{fig:each_match_kawasaki}) compared to the other 17 teams, which implies the aggregated data is still effective to evaluate the robustness of passing networks.

\subsection{Correlation analysis}
The final analysis is to clarify the relationship between the team performances (evaluated by points) and the robustness of the networks (evaluated by $d$ and $S$).
The annual rankings are decided by the points.
The team with the highest points is the champion in the season.
As we showed above, Kawasaki's network was the most robust.
Kawasaki became the champion in the season.
Here, we investigate whether the similar tendency is observed even for the other teams.
Thus, we conducted correlation analyses between the team performance (points) and the robustness of the networks.
We only used the data of node removals for this correlation analysis because it accurately captures the robustness of networks compared to link removals.

\begin{figure}[h]
	\centering
	\includegraphics[width=140mm]{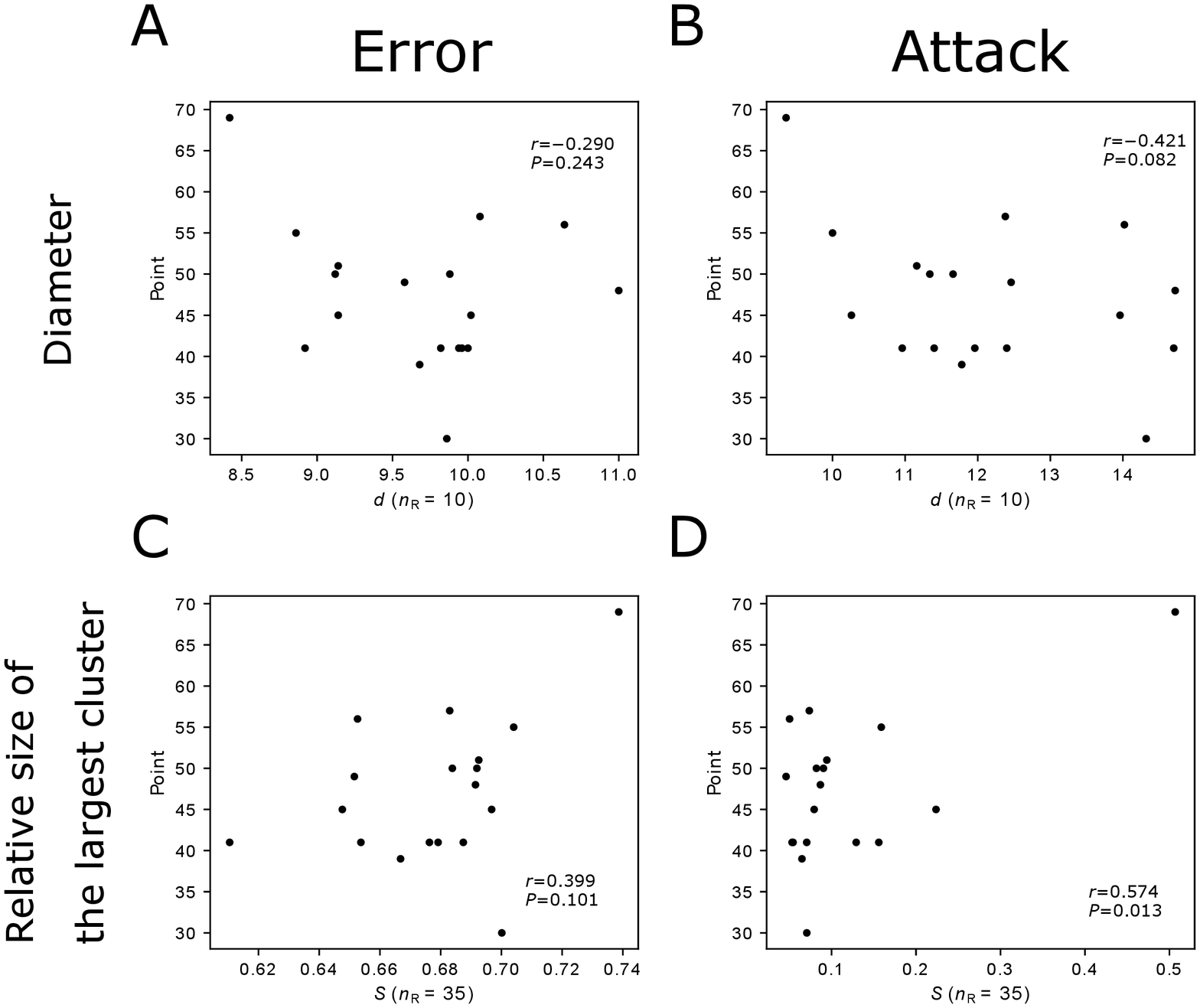}
	\caption{Scatter plots between the points and the diameter $d$ (A and B) and the points and the size of the largest cluster $S$ (C and D), respectively. (A and C) are for errors. (B and D) are for attacks. The Pearson correlation coefficient, $r$, and its $P$ value are also shown.}
	\label{fig:corr_diam}
\end{figure}

Figures \ref{fig:corr_diam}(A) and (B) are the scatter plots between the points and the diameter $d$ at $n_{\rm R}=10$ where (A) corresponds to errors and (B) corresponds to attacks.
We identify whether larger points lead to the robustness of networks (small $d$).
In other words, we check whether a negative correlation is observed.
In the figure, the Pearson correlation coefficient, $r$, and its $P$ value are shown.
As a result, we observed a weak negative correlation in errors ($r=-0.290$) and a negative correlation in attacks ($r=-0.421$), respectively.
However, in both cases, the results are not statistically significant ($P>.05$) although we see a tendency that the diameter is smaller as the points are larger.

Next, we show the scatter plots between the points and the size of the largest cluster $S$ at $n_{\rm R}=35$ for (C) errors and (D) attacks in Fig.~\ref{fig:corr_diam}.
We identify whether larger points lead to the robustness of networks (large $S$).
In other words, we see whether a positive correlation is observed.
The correlation results showed that positive correlations are observed in both errors  ($r=0.399$) and attacks  ($r=0.574$), respectively.
In the case of errors, the result is not statistically significant ($P>.05$).
However, in the case of attacks the result is statistically significant with $P<.05$.
Thus, we can conclude that the size of the largest cluster is higher as the points are larger in the case of attacks.
In short, even if key players become less functional, keeping the connection in a network leads to a win.

\section{Conclusion}
We constructed position-dependent passing networks from 45 matches of all J1 teams.
Then, we continuously removed the nodes or links by two methods, errors or attacks, to analyze the robustness of the networks.

We focused on the change of the diameter $d$, the algebraic connectivity $\lambda_2$, and the size of the largest cluster $S$ to evaluate the robustness of the networks.
The results showed that the passing networks were robust against errors but vulnerable to attacks.
Thus, passing networks are greatly affected by stopping the movement of key players and cutting passes to and from them.

Especially, we found that Kawasaki's network was distinct where the robustness was maintained even if key players making passes were removed. 
There were multiple key players that collect passes in Kawasaki's network.
Kawasaki is known for frequently passing the ball among players.
The style of playing may be similar to tiki-taka used in F.C. Barcelona \cite{Buldu2019SciRep}.

Finally, we conducted the correlation analysis between the points and the diameter $d$ or the points and the size of the largest cluster $S$.
We found that there is a statistically positive correlation between the points and the robustness of the networks in the case of attacks.
We can summarize that the robustness of passing networks is closely tied with the team performance.

Similar studies focused on key players in passing networks where all plays in the matches are included \cite{Grund2012SocNetworks, Pena2012ProcEPSC}.
In contrast, for the first time, by considering continuous node and link removals, our study evaluated the robustness of passing networks after key players or key passes in different locations are removed one after another.
This analysis makes it possible to dynamically evaluate the robustness of passing networks.
Thus, this is a key strength of our study.

There are some limitations in this paper.
Although we adopted an undirected graph for a passing network, a directed graph may be suitable because passes are always to one another.
However, in this study, when a node is removed, links connected to the node are also removed.
Thus, the effect of directions is small for this type of robustness analysis.
From another perspective, football networks are spatio-temporal networks whose structures dynamically change in time and space \cite{Gudmundsson2017ACMComputSurv}.
We partially incorporated spatial information in our position-dependent networks.
However, we ignored the change of dynamical structures of networks in time.
For instance, football passing networks are often different between the first half and the second half.
In another case, a passing network would change after scoring or receiving a goal.
Thus, the dynamical change of networks over time is important.
This can be analyzed by the technique of temporal networks and recent studies used the technique for passing networks \cite{Buldu2019SciRep} or formations \cite{Narizuka2019SciRep} in football games.
Lastly, the importance of passing networks is different depending on the formation \cite{Tamura2015EPJDataSci} or the style of play.
One example is possession versus counter-attack.
The robustness of passing networks may be related to winning in the former compared to the latter.
In this way, there are many interesting problems in football networks which can be tackled by network science.


\section*{Acknowledgment}
We sincerely thank DataStadium Inc., Japan which provided the J1 league data.
Our study is partially supported by the Institute of Statistical Mathematics, Research Center for Medical and Health Data Science. 
We also thank Daiki Miyagawa and Azumi Mamiya for valuable comments on the manuscript.

\section*{Author Contributions}
G.I. designed the study and wrote the manuscript. G.I., T.T., and S.W. performed the analyses.
G.I. and T.T. prepared the figures and interpreted the results.

\section*{Additional Information}
\subsection*{Competing Interests}
The authors declare no competing interests.

\end{document}